\documentclass[aps,twocolumn,showpacs,superscriptaddress]{revtex4}
\begin{document}
\newcommand{\rmd}{{\rm d}}
\newcommand{\rme}{{\rm e}}
\newcommand{\CD}{{\cal D}}
\newcommand{\CQ}{{\cal Q}}
\newcommand{\CE}{{\cal E}}
\newcommand{\CR}{{\cal R}}
\newcommand{\CH}{{\cal H}}
\newcommand{\CM}{{\cal M}}
\newcommand{\CC}{{\cal C}}
\newcommand{\average}[1]{\left\langle #1 \right\rangle_\CD}
\newcommand{\laverage}[1]{\left\langle #1 \right\rangle_{\CD_{\rm \bf i}}}
\newcommand{\gaverage}[1]{\left\langle #1 \right\rangle_{\Sigma}}
\newcommand{\initial}[1]{{#1_{\rm \bf i}}}
\newcommand{\now}[1]{{#1_{\rm \bf 0}}}
\title{Backreaction Issues in Relativistic Cosmology\\
and the Dark Energy Debate\footnote{Lecture provided at the
XII. Brazilian School of Cosmology and Gravitation, Mangaratiba, Rio de Janeiro, Brazil, September 2006.}}
\author{Thomas Buchert}
\affiliation{Arnold Sommerfeld Center for Theoretical Physics,
Ludwig--Maximilians--Universit\"{a}t, Theresienstra{\ss}e 37,
80333 M\"{u}nchen, Germany \\Email: buchert@theorie.physik.uni-muenchen.de}
\affiliation{Department of Mathematics and Applied Mathematics, University of Cape~Town,
Rondebosch 7701, Cape~Town, South Africa}
%
%
\begin{abstract}
The effective evolution of an inhomogeneous universe model in Einstein's theory of gravitation 
may be described 
in terms of spatially averaged scalar variables. This evolution can be modeled by solutions of
a set of Friedmann equations for an effective scale factor, 
with matter and backreaction source terms, where the latter
can be represented by a minimally coupled scalar field  (`morphon field').
We review the basic steps of a description of backreaction effects in relativistic cosmology that lead to refurnishing the standard cosmological equations, but also lay down a number of
unresolved issues in connection with their interpretation within observational cosmology.
\end{abstract}
\maketitle
\tableofcontents
\clearpage
\section{General thoughts: the standard model, the averaging problem and key insights}
\subsection{The standard model of cosmology}
The standard model of cosmology does not, like  the standard model of particle physics, enjoy
appreciable generality; it is based on  
the simplest conceivable class of (homogeneous--isotropic) solutions of Einstein's laws of gravitation.
It is clear that the inhomogeneous properties of the Universe cannot be described by such a strong idealization.
The key issue is whether it can be described so {\it on average}, and this is the subject of considerable
debate and controversy in the recent literature. If the standard model indeed describes the
averaged model, we have to conjecture that backreaction effects, being the main subject of this contribution, are negligible. 
We are striving to discuss most of the related aspects of this debate.

\subsubsection{Dark Energy and Dark Matter}

In the standard model of cosmology one has to conjecture the existence of two constituents,
if observational constraints are met, that both have yet unknown origin: 
first, a dominant repulsive component is thought to exist that
can be modeled either by a positive cosmological constant or a scalar field, e.g. a so--called 
quintessence field, possibly deriving either from 
higher--order Ricci curvature Lagrangians \cite{riccilagrangians}, \cite{capo}
or string--motivated low--energy effective actions \cite{string:review}. 
In any case we should not expect a
{\em fundamental} scalar field to exist in nature, at least not one that can be viewed as a
natural candidate for the relevant effects needed. This expectation is also hampered by the well--known violation of energy conditions of 
a quintessence field that is able to produce late--time volume acceleration of the Universe.
Rather, a scalar field would likely be an {\em effective} one, either stemming from higher--order gravity terms, or effective terms as remnants from higher dimensions that are compactified or even
non--compactified as in brane world cosmologies \cite{maartens:brane}.
As we shall learn below, already classical general relativity allows to
identify effective geometrical terms, simply resulting from inhomogeneities, with an effective scalar field component, the {\em morphon field} \cite{morphon}, a good example of  William of Ockham's {\em razor}.
If we restrict our attention to cosmology and the fitting of extra terms from various different theories to observational data, then those extra terms may also be mapped into morphon fields with different but unambiguously defined physical consequences. 
A review of the status and properties of such models
can be found in  \cite{copeland:darkenergy}, see also \cite{paddy,uzan:darkenergy}. 

Besides this {\em Dark Energy},
there is, secondly, a non--baryonic component that should considerably exceed 
the contribution by luminous and dark baryons and 
massive neutrinos. This {\em Dark Matter} is thought to be provided by exotic forms of matter, 
not yet detected in experiments. According to the {\em concordance model} \cite{lahav}, 
the former comprises almost $3/4$ and the latter almost $1/4$ of the total source of
Friedmann's equations, up to a few percent that have to
be attributed to baryonic matter and neutrinos (in the matter--dominated era).
The intriguing question is whether an explanation of these missing components is the task
of particle physicists, or whether the cosmological model is built on oversimplified priors.

\subsubsection{Uncharted territory beyond the standard model}

The {\em concordance model} is encircled by a large set of observational data that are,
however, orthogonal only within the predefined solution space of a FRW 
(Friedmann--Robertson--Walker) cosmology. This solution space is two--dimensional,
for Friedmann's expansion law derives from Hamilton's constraint restricted to locally
isotropic and hence (by Schur's Lemma) homogeneous distributions of matter and
curvature,
\begin{equation}
\label{hamilton1}
\Omega_m + \Omega_k + \Omega_{\Lambda}\;=\;1\;\;,
\end{equation}
where the standard cosmological parameters are {\em global} and iconized by the
{\em cosmic triangle} \cite{bahcall},
\begin{equation}
\label{cosmictriangle}
\Omega_m := \frac{8\pi G\varrho_H}{3 H^2}\;\;;\;\; \Omega_k := \frac{-k}{a^2 H^2}\;\;;\;\; \Omega_{\Lambda}:= \frac{\Lambda}{3 H^2}\;\;;
\end{equation} 
$\varrho_H (t)$ is the homogeneous matter density, $H(t):={\dot a}/a$ Hubble's function
with the scale factor $a(t)$, $k$ a positive, negative or
vanishing constant related to the three elementary constant--curvature geometries, and
$\Lambda$ is the cosmological constant.

We shall learn below that an extended solution space of an  averaged inhomogeneous universe model is already three--dimensional, when we include inhomogeneities of matter and geometry. Hence, such more realistic models enjoy more parameter freedom, but at the same time
define these parameters physically in terms of volume averages.
How can we be sure that fitting an idealized model, that ignores inhomogeneities, to observational data is not `epicyclic', especially if the model enters as a prior into the process of
interpreting the data?
Confronting observers with the wider class of averaged cosmologies allows them to draw their
data points within a cube of possible solutions {\em and} to differentiate the relevant observational scales reflected by these data; if we `force' them to draw the data points into the
plane of the FRW solutions {\em on every scale}, then they conclude that there are  `dark' components. Thus, we have to exclude that they may have missed something in the projection {\em and} we have to clarify whether the ignorance of scale--dependence of observables in the standard model does not mislead their interpretation. 
Both issues are equally important to judge
the viability of the standard model in observational cosmology: the first is the question of
how backreaction {\em quantitatively} affects the standard cosmological parameters, and 
the second is the comparison of data taken on small scales (e.g. on cluster scales) and data
taken on large scales (e.g. CMB; high--redshift supernovae). Both additional `degrees of
freedom' in interpreting observational data are interlocked in the sense that backreaction
effects may alter the evolution history of cosmological parameters. A comparison of
data taken on different spatial scales has therefore also to be subjected to a critical assessment of data that are taken at different times of the cosmic history:
with backreaction at work, the simple time--scaling of parameters in a FRW cosmology is also lost.

\vspace{3pt}

Before we enter the physics of backreaction that is easy to understand, we have to first
probe some more critical territory in the following subsection.

\subsection{Different `directions' of backreaction}

The notion of {\em averaging} in cosmology is tied to space--plus--time thinking. Despite the success
of general covariance in the four--dimensional formulation of classical relativity, the
cosmologist's way of conceiving the Universe is {\em evolutionary}. This breaking of 
general covariance is in itself an obstacle to appreciating the proper status of cosmological
equations. The standard model of cosmology is employed with the implicit
understanding that there is a global {\em spatial} frame of reference that, if mapped to the
highly isotropic Cosmic Microwave Background, is elevated to a physical frame rather than
a particular choice of a mathematical slicing of spacetime. Restricting attention to an irrotational cosmic continuum of dust (that we shall mostly retain throughout this contribution), the best we can say is that all elements of the cosmic continuum defined by the
homogeneous distribution of matter are in {\em free fall} within that spacetime, and therefore
are preferred relative to eventually accelerating observers with respect to this frame of
reference. Those preferred observers are called {\em fundamental}.
Exploiting the diffeomorphism degrees of freedom we can write the FRW
cosmology in contrived ways, so that nobody would realize it as such. 
This point is raised as a criticism of an averaging framework \cite{wald}, as if this problem
were not there in the standard model of cosmology. Again, the `natural' choice for the matter model `irrotational dust' is a 
collection of {\em free--falling} continuum elements, now for an inhomogeneous continuum.

Let us consider the 4--metric form to be
\begin{equation}
\label{fourmetric}
^4{\bf g} = -dt^2 + \;^3{\bf g}\;\;;\;\;^3{\bf g}=g_{ab}\,dX^a \otimes dX^b \;\;,
\end{equation}
where latin indices run through $1 \cdots 3$ and $X^a$ are local (Gaussian normal)
coordinates. Evolving the first fundamental form $^3{\bf g}$ of the spatial hypersurfaces
along $\partial / \partial t =:\partial_t$ defines their second fundamental form
\begin{equation}
\label{FF2}
^3{\bf K} = K_{ab}\,dX^a \otimes dX^b \;\;;\;\;K_{ab}:= -\frac{1}{2}\partial_t {g}_{ab}\;\;,
\end{equation}
with the extrinsic curvature components $K_{ab}$.
Such a {\em comoving (synchronous)} slicing of spacetime may be considered `natural', but it may also be
questioned. However, to dismiss its physical relevance due to the fact that shell--crossing
singularities arise is shortsighted. It is a problem of the {\em matter model} in the first place.
A comoving (Lagrangian) frame helps to access nonlinear stages of structure evolution,
as is well--exemplified in Newtonian models of structure formation, where the `slicing
problem' is absent. Those nonlinear stages
inevitably include the development of singularities, provided we do not improve on the matter model
to include forces that counteract gravity (like velocity dispersion) in order to regularize such 
singularities \cite{adhesive}. 
If a chosen slicing appears to be better suited, because it does not run into
singularities, then one should rather ask the question whether the evolution of variables
is restricted to a singular--free regime just because inhomogeneities are not allowed to enter nonlinear stages of structure evolution. An example for this is perturbation theory formulated e.g. in longitudinal gauge, where the variables are `gauge--fixed' to a (up to a given time--dependent scale factor)
non--evolving background. 

However, the {\em slicing problem} is not off the table. There exist strategies to consolidate the notion
of an effective spatial slicing that would minimize frame fluctuations being attributed
to the diffeomorphism degrees of freedom in an inhomogeneous model. Such, more
involved, strategies 
relate to the  {\em intrinsic direction} of backreaction that we put into perspective below.

\subsubsection{Extrinsic (kinematical) and intrinsic backreaction}

Having chosen a {\em foliation of spacetime} implies that we can speak of two `directions':
one being {\em extrinsic} in the direction of the extrinsic curvature $K_{ab}$ of the embedding of the hypersurface into spacetime (e.g. parametrized by time),
the other being {\em intrinsic} in the direction of the Ricci tensor $R_{ab}$ of the three-dimensional
spatial hypersurfaces parametrized by a scaling parameter (let it be the geodesic radius of a randomly placed geodesic ball). 
Consequently, we may speak of two `directions' of
backreaction: inhomogeneities in extrinsic curvature and in intrinsic curvature.
The former is of {\em kinematical nature}, since we may interpret the extrinsic
curvature actively through the expansion tensor 
$\Theta_{ab}:=-K_{ab}$, and introduce a split into its kinematical parts: 
$\Theta_{ab} = \frac{1}{3}g_{ab}\Theta + \sigma_{ab}$, with the rate of expansion 
$\Theta =\Theta^c_{\;c}$, the shear tensor $\sigma_{ab}$, and the rate of shear
$\sigma^2 :=1/2 \sigma_{ab}\sigma^{ab}$;
note that vorticity and acceleration are absent for dust in the present  flow--orthogonal foliation. The latter addresses the so--called {\em fitting problem} \cite{ellis,ellisstoeger},
\cite{singh2}, i.e. the question whether
we could find an effective constant--curvature geometry that replaces the inhomogeneous
hypersurface. An answer to this question has to deal with the problem of `averaging' the tensorial geometry
for which several different strategies are conceivable. Some of those strategies do not distinguish between extrinsic and intrinsic averaging (e.g. \cite{zala97}, \cite{zala05}, and other 
references in  \cite{ellisbuchert}). 
One method has recently
obtained a strong position in the context of Perelman's work (e.g. \cite{perelman:entropy}) on the Ricci--Hamilton flow
related to the recent proof of Poincar\'e's conjecture, and implied progress on Thurston's geometrization program to cut a Riemannian manifold into `nice pieces' of eight
elementary geometries. This method we briefly scetch now.

\subsubsection{Smoothing the geometry}
\label{subsubsect:ricciflow}

Employing the Ricci--Hamilton flow \cite{hamilton:ricciflow1}, \cite{hamilton:ricciflow2},
an `averaging' of geometry can be put into practice by a rescaling of the metric tensor. 
A general scaling flow is described by Petersen's equations \cite{petersen} that we may implement through
a 2+1 setting by evolving the boundary of a geodesic ball in a three--dimensional
cosmological hypersurface in radial directions, thus exploring the Riemannian manifold
passively. Upon linearizing the general scaling flow, e.g. in normal geodesic coordinates,
we obtain a scaling equation for the metric along radial directions \cite{klingon};
up to tangential geometrical terms on the boundary we obtain:
\begin{equation}
\frac{\partial}{\partial r}\,g_{ab}(r) - \frac{\partial}{\partial r}\,g_{ab}(r)\Big\vert_{r_0} = -2 {\cal R}_{ab} (r_0 ) [r-r_0 ]\;,
\end{equation}
i.e. the metric scales in the direction of its Ricci tensor much in the same way as it is deformed in the direction of the extrinsic curvature by the Einstein flow.
If we now implement the active (geometrically Lagrangian) point of view of deforming
the metric by the same flow along a Lagrangian vector field $\partial / \partial r_0$ while holding
the geodesic radius $r_0$ fixed, we are able to smooth the metric in a controlled way. Depending
on our choice of normalization of the flow, we may preserve the mass content inside
the geodesic ball while smoothing the metric.
Such a rescaling transforms average properties on given hypersurfaces from their values
in the inhomogeneous geometry (the actual space section) to their values on a constant--curvature geometry (the fitting template for the space section): they are
{\em renormalized} resulting in additional backreaction effects due to the difference of the
two volumes (the Riemannian volume of the actual space section and the constant--curvature
volume) -- the {\em volume effect}, and also {\em curvature backreaction} terms that involve averaged invariants
of the Ricci tensor. For details and references see \cite{klingon} and for small overviews 
\cite{buchertcarfora:PRL} and \cite{buchertcarfora3}.
In such a setting the role of lapse and shift functions (i.e. the slicing problem) can also be controlled by employing the recent results of Perelman \cite{carforabuchert:perelman}.

We now come to some crucial points of understanding the physics behind backreaction.
In order not to think of any exotic mechanism, the historical use of the notion `model with backreaction' should simply be replaced by `more realistic model'.

\subsection{The origin of kinematical backreaction and the physics behind it}

Let us now concentrate on the question, {\em why} there must be backreaction at work,
restricting attention to {\em kinematical backreaction} as defined above. After
we have understood the reasons behind backreaction effects in general terms, i.e. without resorting
to restrictions of spatial symmetry or approximations of evolution models, the very question of their relevance is better defined.

\subsubsection{An incomplete message to particle physicists}

Employing Einstein's general theory of relativity to describe the evolution of 
the Universe, we base our universe model on a relation between geometry and matter sources. A maximal reduction
of this theoretical fundament is to consider the simplest conceivable geometry. 
Without putting in doubt that it might be an oversimplification to assume a locally isotropic
(and hence homogeneous) geometry, standard cosmology conjectures the existence of sources that would generate this simple geometry. As already remarked, the majority of these
sources have yet unknown physical origin. Obviously, particle physicists take
the demand for missing fundamental fields literally. But, as was emphasized above, 
the standard model has physical sense only, if a homogeneous--isotropic solution of
Einstein's equations also describes
the inhomogeneous Universe effectively, i.e. {\em on average}.
This is not obvious. The very fact that the distributions of matter and geometry are
inhomogeneous gives rise to backreaction terms; we shall restrict them to those  additional terms that influence the kinematics of the homogeneous--isotropic solutions.
These terms can be viewed to arise on the geometrical side of Einstein's equations, but they
may as well be put on the side of the sources. 

We start with a basic kinematical observation that lies at the heart of the backreaction problem.

\subsubsection{Non--commutativity}

Let us define spatial averaging of a scalar field $\Psi$ on a compact spatial domain $\CD$ 
with volume $V_\CD : = \vert \CD \vert$ through its Riemannian volume average
\begin{equation}
\label{averager}
\average{\Psi (X^i ,t)}: = 
\frac{1}{V_\CD}\int_\CD  \Psi (X^i, t) \,J d^3 X\;\,;\;\,J:=\sqrt{\det(g_{ij})}\;.
\end{equation}
The key property of inhomogeneity of the field 
$\Psi$ is revealed by the {\em commutation rule} \cite{buchertehlers}, \cite{buchert:grgdust}:
\begin{equation}
\label{commutationrule}
\partial_t \langle\Psi \rangle_\CD - 
\langle\partial_t \Psi \rangle_\CD 
= \average{\Theta\Psi} - \average{\Theta}\average{\Psi}\;,
\end{equation}
where $\Theta$ denotes the trace of the expansion tensor; its average describes the rate of volume change of
a collection of fluid elements along $\partial /\partial t$,
\begin{equation}
\label{effectivehubble}
 \average{\Theta} = \frac{\partial_t V_\CD}{V_\CD} = : 3H_\CD \;,
\end{equation} 
where we have introduced a volume Hubble rate $H_\CD$ that reduces to Hubble's function
in the homogeneous case.
{\em Commutativity} reflects the conjecture of the standard model which implies that a realistically evolved inhomogeneous field will feature the same average characteristics as those predicted by the evolution
of the (homogeneous) average quantity; in other words, the right--hand--side of 
(\ref{commutationrule}) is assumed to vanish. This rule also shows that backreaction terms
deal with the sources of {\em non--commutativity} that are in general non--zero for inhomogeneous fields. 

\subsubsection{Regional volume acceleration despite local deceleration}

Based on a first application of the above rule, we shall 
emphasize that there is not necessarily {\em anti--gravity} at work, e.g. in
the `redcapped' version of a positive cosmological constant, in order to have sources
that counteract gravity. Raychaudhuri's equation, if physically essential terms like
vorticity, velocity dispersion, or pressure are retained, provides terms needed 
to oppose gravity, e.g., to support spiral galaxies (vorticity), elliptical galaxies (velocity
dispersion), and other stabilization mechanisms involving pressure (think of the hierarchy of stable states of stars
until they collapse into a Black Hole). Admittedly, those terms
are effectively `small--scale--players'. Now, let us consider Raychaudhuri's equation restricted
to irrotational dust,
\begin{equation}
\label{localraychaudhuri}
\partial_t \Theta = \Lambda - 4\pi G \varrho + 2{\rm II} - {\rm I}^2 \;\;,
\end{equation}
with the principal scalar invariants of $\Theta_{ab}$, $2{\rm II}:= 2/3 \Theta^2 - 2\sigma^2$ and ${\rm I}:=\Theta$.
Then, unless there is a positive cosmological constant, there is no term that could 
counter--balance gravitational attraction and, at every point, ${\partial_t \Theta} <0$.
Applying the {\em commutation rule} ({\ref{commutationrule}) for $\Psi = \Theta$,
we find that the averaged variables obey the same equation as above {\em despite}
non--commutativity \footnote{This is
only true, if all terms appearing in Raychaudhuri's equation are principal scalar invariants;
it is actually a special non--linearity of this equation that cancels the corresponding non--commutativity
term (see Corollary I in \cite{buchert:grgdust}).}:
\begin{equation}
\label{regionalraychaudhuri}
\partial_t \average{\Theta} = \Lambda - 4\pi G \average{\varrho} + 2\average{{\rm II}} - 
\average{\rm I}^2\;\;.  
\end{equation}
This result can be understood on the grounds that shrinking the domain $\CD$ to a point
should produce the corresponding local equation. Now, notwithstanding, the above equation
contains a positive term that acts against gravity. This can be easily seen by rewriting the
averaged principal invariants: we obtain $2\average{\rm II} - \average{\rm I}^2 = 
2/3 \average{(\Theta - \average{\Theta})^2} - 1/3 \average{\Theta}^2 - 2\average{\sigma^2}$ which, compared with
the corresponding local term $2{\rm II}- {\rm I}^2 = -1/3 \Theta^2 -2\sigma^2$, gave rise to a
positive--definite averaged variance.
It may appear
`magic' that the time--derivative of a (on some spatial domain $\CD$) averaged expansion may be positive despite the time--derivative of the expansion {\em at all points} in $\CD$ is negative. 
As the above explicit calculation shows, this is not `unphysical' \cite{wald} but simply due to the fact that
an average correlates the local contributions, and it is this correlation (or fluctuation) that
acts in the sense of a `kinematical pressure'. The interesting point is that this positive term
is a `large--scale player', as we shall make more precise below (the physical and observational consequences
of this term have been thoroughly explained and illustrated in the review paper \cite{rasanen}).

What we should learn from this simple exercise is that any {\em local} argument, e.g.
on the smallness of some perturbation amplitude at a given point, is not enough to exclude 
{\em regional} (`global') physical effects that arise from averaging inhomogeneities. As we shall also learn below, such
correlation effects must not be subdominant compared to the magnitude of the local fields,
since they are related to the spatial variation of the local fields and, having said `spatial', it could
(and it will) imply a coupling to the geometry as a dynamical variable in Einstein gravitation. 
This latter remark will turn out very useful in understanding the potential relevance of backreaction
effects in relativistic cosmology.

\subsubsection{The production of information in the Universe}
\label{subsect:information}

The above considerations on effective expansion properties can be essentially traced
back to `non--commutativity' of averaging and time--evolution, lying at the root of
backreaction. (Note that additional backreaction terms that have been discussed in
Subsection~\ref{subsubsect:ricciflow} are also the result of a `non--commutativity', this time
between averaging and spatial rescaling -- see also \cite{ellisbuchert}.) 
The same reasoning applies
to the `production of information' in the following sense. Applying the {\em commutation rule}
(\ref{commutationrule}) to the density field, $\Psi = \varrho$,
\begin{equation}
\label{commutationentropy}
\average{\partial_t \varrho}-\partial_t \average{\varrho}\; =\;
\frac{{\partial_t \,{\cal S}\lbrace\varrho || \average{\varrho}\rbrace}}
{V_{\CD}}\;\;,
\end{equation}
we derive, as a source of non--commutativity, 
the (for positive--definite density) positive--definite Lyapunov functional (known as {\it Kullback--Leibler functional} in information theory; \cite{hosoya:infoentropy} and 
references therein):
\begin{equation}
\label{relativeentropy}
{\cal S} \lbrace \varrho || \average{\varrho}\rbrace: \;=\; 
\int_{\CD} \varrho \ln \frac{\varrho}{\average{\varrho}}\,J d^3 X \;\;.
\end{equation}
This measure vanishes for Friedmannian cosmologies (`zero structure'). 
It attains some {\em positive} time--dependent value otherwise.
The source in (\ref{commutationentropy}) shows that relative entropy production and
volume evolution are competing:  commutativity can be reached, if the volume expansion
is faster than the production of information contained within the same volume.
 
In \cite{hosoya:infoentropy} the following conjecture was advanced:

\smallskip

{\sl The relative information entropy of a dust matter model
${\cal S} \lbrace \varrho || \gaverage{\varrho}\rbrace$
is, for sufficiently large times, globally (i.e. averaged over the whole
manifold $\Sigma$ that is assumed simply--connected and without boundary) an increasing function of time.}

\medskip

\noindent
This conjecture already holds for linearized scalar perturbations on a Friedmannian 
background (the growing--mode
solution of the linear theory of gravitational instability implies $\partial_t \,{\cal S} >0$
and $\cal S$ is, in general, time--convex, i.e. $\partial_t^2 \,{\cal S} >0$). 
Generally, information entropy is produced, i.e. $\partial_t \,{\cal S} >0$ with
\begin{equation}
\label{entropyproduction}
\frac{{\partial_t \,{\cal S}\lbrace\varrho || \average{\varrho}\rbrace}}
{V_{\CD}} = -\average{\delta\varrho \Theta} = 
-\average{\varrho \delta\Theta} = -\average{\delta\varrho \delta\Theta}\;, 
\end{equation}
(and with the deviations of the local fields from their average values, e.g.  $\delta\varrho :=
\varrho - \average{\varrho}$), 
if the domain
$\CD$ contains more {\em expanding underdense} and {\em contracting overdense} regions than the opposite states 
{\em contracting underdense} and {\em expanding overdense} regions. 
The former states are clearly favoured in the course of evolution,
as can be seen in simulations of large--scale structure. 

There are essentially three lessons relevant to the origin of backreaction that can be learned here.
First, structure formation (or `information' contained in structures) installs a positive--definite
functional as a potential to increase the deviations from commutativity; it can therefore
not be statistically `averaged away' (the same remark applies to the averaged variance of the expansion rate discussed before). 
Second, gravitational instability acts in the form of a negative feedback that enhances 
structure (or `information'), i.e. it favours contracting clusters and expanding voids. This tendency 
is opposite to the thermodynamical interpretation within a closed system where such a 
relative entropy would decrease and the system would tend to thermodynamical equilibrium.
This is a result of the long--ranged nature of gravity: the system contained within $\CD$ must
be treated as an open system.  Third, backreaction is a genuinly 
non--equilibrium phenomenon, thus, opening this subject also to the language of non--equilibrium thermodynamics \cite{prigogine}, \cite{schwarz}, \cite{zimdahl}. `Near--equilibrium' can only be maintained (not established) by a 
simultaneous strong volume expansion of the system. 
Later we discuss an example of a cosmos that is `out--of--equilibrium', i.e. settled in a state
far from a Friedmannian model that can be associated with the relative equilibrium state 
${\cal S} = 0$. 

In particular, we conclude that the standard model may be a good description for the averaged
variables only when information entropy production is {\em over}--compensated by volume expansion,
a property that is realized by linear perturbations on a FRW background. Thus, the question
is whether this remains true in the nonlinear regime, where information production is
expected to be more efficient.

Before we can go deeper into the problem of whether such backreaction terms, being
well--motivated, are indeed relevant in a {\em quantitative} sense, we have to study the governing
equations.

\section{Refurnishing the cosmological equations}

In this section we recall a set of averaged Einstein equations together with alternative
forms of these equations which put us in the position to study backreaction terms as
additional sources to the standard Friedmann equations.

\subsection{Averaged cosmological equations} 

In order to find evolution equations for effective (i.e. spatially averaged) cosmological
variables, we may put the following simple idea into practice. We observe that 
Friedmann's differential equations capture the scalar parts of Einstein's equations, 
while restricting them by the strong symmetry assumption of local isotropy.
The resulting equations,
Friedmann's expansion law (the energy or Hamiltonian constraint)  and Friedmann's acceleration law (Raychaudhuri's equation),
\begin{equation}
\label{friedmann1}
3\left(\frac{\dot a}{a}\right)^2 - 8\pi G\varrho_H  - \Lambda \;=\;-\frac{3k}{a^2}\;\;;
\end{equation}
\begin{equation}
\label{friedmann2}
3\frac{\ddot a}{a} + 4\pi G\varrho_H - \Lambda\;=\;0\;\;,
\end{equation}
can be replaced by their spatially averaged, {\em general} counterparts
(for the details
the reader is referred to \cite{buchert:grgdust,buchert:grgfluid,buchert:static,morphon};
an overdot abbreviates, in both cases, the covariant time--derivative $\partial_t = 
u^{\mu}\partial_{\mu}$):
\begin{equation}
\label{averagedhamilton}
3\left( \frac{{\dot a}_\CD}{a_\CD}\right)^2 - 8\pi G \average{\varrho}-\Lambda \;=\; - \frac{\average{\CR}+{\cal Q}_\CD }{2} \;\;;
\end{equation}
\begin{equation}
\label{averagedraychaudhuri}
3\frac{{\ddot a}_\CD}{a_\CD} + 4\pi G \average{\varrho} -\Lambda\;=\; {\cal Q}_\CD\;\;.
\end{equation}
We have replaced the Friedmannian scale factor by the {\em volume scale factor} $a_\CD$, depending  on content,  shape and position of the domain of averaging $\CD$,  defined via the
domain's volume  $V_{\CD}(t)=|\CD|$, and the initial 
volume $V_{\initial\CD}=V_{\CD}(\initial{t})=|\initial{\CD}|$:
\begin{equation}
\label{volumescalefactor}
a_\CD (t) := \left( \frac{V_{\CD}(t)}{V_{\initial\CD}} \right)^{1/3} \;\;.
\end{equation}
Using a scale factor instead of the volume should not be confused with `isotropy'. The above
equations are general for the evolution of a mass--preserving, compact domain containing an 
irrotational continuum of dust, i.e. they
provide a background--free and non--perturbative description of inhomogeneous {\em and} anisotropic fields. (One could, of course, introduce an isotropic or anisotropic  reference background  \cite{buchertehlers} or, explicitly isolate an averaged shear from the above
equations to study deviations
from the  kinematics of Bianchi--type models, as was done with some interesting results
in \cite{barrowtsagas}.)   
The new term appearing in these equations, the {\em kinematical backreaction}, arises as
 a result of expansion and shear fluctuations:
\begin{equation}
\label{backreactionterm} 
{\cal Q}_\CD : = 2 \average{\rm II} - \frac{2}{3}\average{\rm I}^2 =
\frac{2}{3}\average{\left(\theta - \average{\theta}\right)^2 } - 
2\average{\sigma^2} \;;
\end{equation}
$\rm I$  and $\rm II$  denote  the principal scalar invariants  of the  extrinsic
curvature, and the second equality  follows  by introducing the
decomposition
of the  extrinsic curvature into  the kinematical variables, as before. 
Also, it is not a surprise that the general averaged 3--Ricci curvature $\average{\cal R}$ replaces the constant--curvature term in Friedmann's equations. 
Note also that the term $\CQ_\CD$ encoding the fluctuations has the particular structure of
vanishing
at a Friedmannian background, a property that it shares with gauge--invariant variables.

In the Friedmannian case, Eq.~(\ref{friedmann2}) arises as the time--derivative
of Eq.~(\ref{friedmann1}), if the integrability condition of restmass--conservation is respected,
i.e. the homogeneous density $\varrho_H \propto a^{-3}$. In the general case, however, 
restmass--conservation is not sufficient. In addition to the (built--in) integral 
\begin{equation}
\label{averageddensity}
\average{\varrho} = \frac{\laverage{\varrho(\initial{t})}}{a_\CD^3} 
= \frac{M_\CD}{a_\CD^3 V_{\initial\CD}} \;\;;\;\;M_\CD = M_{\initial\CD}\;\;,
\end{equation}
we also have to respect the following {\em curvature--fluctuation--coupling}:
\begin{equation}
\label{integrability}
\frac{1}{a_\CD^6}\partial_t \left(\,{\cal Q}_\CD \,a_\CD^6 \,\right) 
\;+\; \frac{1}{a_\CD^{2}} \;\partial_t \left(\,\average{\CR}a_\CD^2 \,
\right)\,=0\;.
\end{equation}
This relation will be key to the understanding of how backreaction can take the role
of {\em Dark Energy}.

\subsection{Alternative forms of the averaged equations}

We here provide three compact forms of the averaged equations introduced above,
as well as some derived quantities.  They will prove useful for our
further discussion of the backreaction problem.

\subsubsection{Generalized expansion law}

The correspondence  between Friedmann's expansion law (\ref{friedmann1}) and the
general expansion law (\ref{averagedhamilton}) can be made more explicit through formal integration of the integrability condition (\ref{integrability}):
\begin{equation}
\label{integrabilityintegral}
\frac{3k_{\initial\CD}}{a_\CD^2} - \frac{1}{ a_\CD^2} \int_{t_0}^t \,dt' \;
{\cal Q}_\CD\; \frac{d}{dt'} a^2_\CD (t')
= \frac{1}{2}\left(\langle {\cal R} \rangle_\CD + {\cal Q}_\CD\right) \;\;. 
\end{equation}
The (domain--dependent) integration constant $k_{\initial\CD}$ relates the new terms to
the `constant--curvature part'. We insert this latter integral back into the expansion law
(\ref{averagedhamilton}) and obtain:
\begin{equation}
\label{generalexpansionlaw}
3\frac{\dot{a}_\CD^2 + k_{\initial\CD}}{a_\CD^2 } - 8\pi G \average{\varrho}
- \Lambda = \frac{1}{ a_\CD^2} \int_{\initial{t}}^t \rmd t'\ {\cal Q}_\CD
\frac{\rmd }{\rmd t'} a^2_\CD(t')\;\;.
\end{equation}
This equation is formally equivalent to its Newtonian counterpart \cite{buchertehlers}.
It shows that, by eliminating the averaged scalar curvature, the whole history of the averaged 
kinematical fluctuations acts as a source of a generalized expansion law that features the
`Friedmannian part' on the left--hand--side of (\ref{generalexpansionlaw}).

\subsubsection{Effective Friedmannian framework}

We may also recast the general equations (\ref{averagedhamilton}, \ref{averagedraychaudhuri}) by appealing to the Friedmannian framework.
This amounts to reinterpret geometrical terms, that arise through averaging, as 
effective sources within a Friedmannian setting.
In the present case the averaged equations can be written as standard zero--curvature
Friedmann equations for an {\em effective perfect fluid energy momentum tensor}
with new effective sources \cite{buchert:grgfluid}:
\begin{eqnarray}
\label{equationofstate}
\varrho^{\CD}_{\rm eff} = \average{\varrho}-\frac{1}{16\pi G}{\cal Q}_\CD - 
\frac{1}{16\pi G}\average{\CR}\;\;;\nonumber\\
{p}^{\CD}_{\rm eff} =  -\frac{1}{16\pi G}{\cal Q}_\CD + \frac{1}{48\pi
G}\average{\CR}\;\;.
\end{eqnarray}
\begin{eqnarray}
\label{effectivefriedmann}
3\left(\frac{{\dot a}_\CD}{a_\CD}\right)^2 - 8\pi G \varrho^{\CD}_{\rm eff} - \Lambda\;=\;0\;\;;\nonumber\\
3\frac{{\ddot a}_\CD}{a_\CD}  +  4\pi G (\varrho^{\CD}_{\rm eff}
+3{p}^{\CD}_{\rm eff}) - \Lambda \;=\;0\;\;;\nonumber\\
{\dot\varrho}^{\CD}_{\rm eff} + 
3 \frac{{\dot a}_\CD}{a_\CD} \left(\varrho^{\CD}_{\rm eff}
+{p}^{\CD}_{\rm eff} \right)\;=\;0\;\;.
\end{eqnarray}
Eqs.~(\ref{effectivefriedmann}) correspond to the equations
(\ref{averagedhamilton}),
(\ref{averagedraychaudhuri}) and (\ref{integrability}), respectively. 
(Note that we may retain the constant--curvature part in the (scale--dependent) 
Friedmannian expansion law and instead consider only the deviations of averaged curvature from constant curvature as sources (see \cite{buchert:static}).)

\subsubsection{`Morphed' Friedmann cosmologies}
\label{subsect:morphon}

In the above--introduced framework we distinguish the averaged matter source on the one hand, and averaged 
sources due to geo\-metrical inhomogeneities stemming from extrinsic and intrinsic curvature
(kinematical backreaction terms) on the other. As shown above, 
the averaged equations can be written as standard Friedmann equations that are sourced 
by both. Thus, we have the choice to consider the averaged model as a (scale--dependent) `standard model' with matter source evolving in a {\em mean field} of backreaction terms.
This form of the equations is closest to the standard model of cosmology. It is a `morphed' 
Friedmann cosmology, sourced by matter and `morphed' by a (minimally coupled) scalar field, the {\em morphon field} \cite{morphon}. We write
(recall that we have no matter pressure source
here):
\begin{equation}
\label{morphon:sources}
\varrho^\CD_{\rm eff} =: \langle\varrho\rangle_{\cal D} +
\varrho^\CD_{\Phi}\;\;\;;\;\;\;
p^\CD_{\rm eff} =: p^\CD_{\Phi}\;\;,
\end{equation}
with 
\begin{equation}
\label{morphon:field}
\varrho^\CD_{\Phi}=\epsilon \frac{1}{2}{{\dot\Phi}_\CD}^2 + U_\CD\;\;\;;\;\;\;p^\CD_{\Phi} =
\epsilon \frac{1}{2}{{\dot\Phi}_\CD}^2 - U_\CD\;\;,
\end{equation}
where $\epsilon=+1$ for a standard scalar field (with positive kinetic energy), and 
$\epsilon=-1$ for a phantom scalar field (with negative kinetic energy);
(we have chosen the letter $U$ for the potential
to avoid confusion with the volume functional; $\epsilon$ arises if we insist on a
real--valued scalar field).
Thus, in view of Eq.~(\ref{equationofstate}), we obtain the following correspondence:
\begin{equation}
\label{correspondence1}
-\frac{1}{8\pi G}{\cal Q}_\CD \;=\; \epsilon {\dot\Phi}^2_\CD - U_\CD\;\;\;;\;\;\;
-\frac{1}{8\pi G}\average{\CR}= 3 U_\CD\;\;.
\end{equation} 
Inserting (\ref{correspondence1}) into the integrability condition (\ref{integrability}) 
then implies that $\Phi_\CD$, for 
${\dot\Phi}_\CD \ne 0$, obeys the (scale--dependent) Klein--Gordon equation:
\begin{equation}
\label{kleingordon}
{\ddot\Phi}_\CD + 3 H_{\cal D}{\dot\Phi}_\CD + 
\epsilon\frac{\partial}{\partial \Phi_\CD}U(\Phi_\CD)\;=\;0\;\;.
\end{equation} 
This correspondence allows us to interpret the kinematical backreaction effects in terms
of properties of scalar field cosmologies, notably quintessence or phantom--quintessence
scenarii that are here routed back to models of inhomogeneities. 
For a full--scale discussion of this correspondence see \cite{morphon}.

\subsubsection{A note on closure assumptions}

This system of the averaged equations in the various forms introduced above does not close unless we specify a model for the inhomogeneities.
Note that, if the system would close, this would mean that we solved the scalar parts of the
GR equations in
general by reducing them to a set of ordinary differential equations
on arbitrary scales. 
Closure assumptions have been studied by prescribing a {\it cosmic equation of state}
of the form 
$p^\CD_{\rm eff} = \beta (\varrho^\CD_{\rm eff}, a_{\cal D})$ \cite{buchert:darkenergy,
buchert:static}, or by prescribing the backreaction terms through {\em scaling solutions}, e.g.
${\cal Q}_\CD \propto a_\CD^n$ \cite{morphon}.
We shall come back to the important question of how to close the averaged equations in Subsect.~\ref{subsect:approximations}.

\subsection{Derived dimensionless quantities}

For any quantitative discussion it is important to provide a set of dimensionless characteristics
that arise from the above framework.  

\subsubsection{The cosmic quartet}

We start by dividing the volume--averaged Hamiltonian constraint (\ref{averagedhamilton})
by the squared {\em volume Hubble functional} $H_\CD := {\dot a}_\CD / a_\CD$ introduced before.
Then, expressed through the following set of `parameters' \footnote{We shall, henceforth, call these characteristics `parameters', but the reader should keep in 
mind that these are  functionals on $\CD$.},
\begin{eqnarray}
\label{omega}
\Omega_m^{\CD} : = \frac{8\pi G}{3 H_{\CD}^2} \langle\varrho\rangle_{\cal D}  \;\;;\;\;
\Omega_{\Lambda}^{\CD} := \frac{\Lambda}{3 H_{\CD}^2 }\;\;;\nonumber\\
\Omega_{\cal R}^{\CD} := - \frac{\average{\cal R}}{6 H_{\CD}^2 }\;\;;\;\;
\Omega_{\cal Q}^{\CD} := - \frac{{\cal Q}_{\CD}}{6 H_{\CD}^2 } \;\;,
\end{eqnarray}
the averaged Hamiltonian constraint assumes the form of a {\it cosmic quartet} 
\cite{buchert:jgrg,buchertcarfora3}:
\begin{equation}
\label{hamiltonomega}
\Omega_m^{\CD}\;+\;\Omega_{\Lambda}^{\CD}\;+\;\Omega_{\cal R}^{\CD}\;+\;
\Omega_{\cal Q}^{\CD}\;=\;1\;\;,
\end{equation}
showing that the solution space of an averaged inhomogeneous cosmology is three--dimensional in the present framework.
In this set, the averaged scalar curvature parameter and the kinematical backreaction parameter
are directly expressed through $\average{\cal R}$ and ${\cal Q}_{\CD}$, respectively.
In order to compare this pair of parameters with the `constant--curvature parameter' 
that is the only curvature contribution
in standard cosmology to interpret
observational data, we can alternatively introduce the pair
\begin{equation}
\label{omeganewton}
\Omega_{k}^{\CD} := - \frac{k_{\initial\CD}}{a_\CD^2 H_{\CD}^2 }\;;\;
\Omega_{{\cal Q}N}^{\CD} := \frac{1}{3 a_\CD^2 H_\CD^2}
\int_{\initial{t}}^t \rmd t'\ {\cal Q}_\CD\frac{\rmd }{\rmd t'} a^2_\CD(t')\;,
\end{equation}
being related to the previous parameters by
\begin{equation}
\label{parameterrelation}
\Omega_{k}^{\CD} +\Omega_{{\cal Q}N}^{\CD}\;=\; 
\Omega_{\cal R}^{\CD} + \Omega_{\cal Q}^{\CD}\;\;.
\end{equation}
After a little thought we see that both sides of this equality would mimick a cosmological
`constant' in a Friedmannian model. Note, in particular, that it is not the additional backreaction parameter alone that can play this role, but it is the joint action with the
(total) curvature parameter, or, looking to the left--hand--side, it is the cumulative effect acquired during the history of the backreaction parameter. A positive cosmological constant would require this sum, or the effective history, respectively, to be positive.

\subsubsection{Volume state finders}

Like the volume scale factor $a_\CD$ and the volume Hubble functional $H_\CD$, we may 
introduce `parameters' for higher derivatives of the volume scale factor, e.g. the {\em volume deceleration
functional}
\begin{equation}
\label{deceleration}
q^\CD := -\frac{{\ddot a}_\CD}{a_\CD}\frac{1}{H_\CD^2} = \frac{1}{2}
\Omega_m^{\CD} + 2 \Omega_{\cal Q}^{\CD} - \Omega_{\Lambda}^{\CD}\;\;.
\end{equation}
Following  \cite{sahnietal,alametal} (see also \cite{evans} and  
references therein) we may also define the following
{\it volume state finders} involving the third derivative of the volume scale factor:
\begin{equation}
\label{statefinder1}
r^\CD := \frac{{\dddot a}_\CD}{a_\CD}\frac{1}{H_\CD^2} = 
\Omega_m^{\CD}(1 + 2 \Omega_{\cal Q}^{\CD}) + 2 \Omega_{\cal Q}^{\CD}
(1+4\Omega_{\cal Q}^{\CD}) - \frac{2}{H_\CD}{\dot\Omega}_{\cal Q}^{\CD} \;\;,
\end{equation}
and
\begin{equation}
\label{statefinder2}
s^\CD := \frac{r^\CD -1}{3(q^\CD - 1/2)}\;\;. 
\end{equation}
The above definitions are identical to those given in  \cite{sahnietal,alametal}, 
however, note the following obvious and subtle differences.
One of the obvious differences was already mentioned: while the usual state finders of 
a global homogeneous state in the 
standard model of cosmology are the same for every scale, the volume state finders defined
above are different for different scales. The other is the fact that the volume state finders 
apply to an inhomogeneous cosmology with arbitrary 3--metric, while the usual state finders
are restricted to a FRW metric.  Besides these there is a more subtle difference, namely a degeneracy in the Dark Energy density parameter:
while \cite{sahnietal,alametal} denote (with obvious adaptation) 
$1-\Omega_m^\CD = \Omega_{X}^\CD$ we have from Hamilton's constraint 
(\ref{hamiltonomega}) $\Omega_{X}^\CD = \Omega_{\cal Q}^{\CD} + \Omega_{\cal R}^\CD$, i.e. so--called $X$--matter is composed of two physically distinct components.

\subsubsection{Cosmic equation of state and Dark Energy equation of state}

We already mentioned the possibility to characterize a solution of the averaged
equations by a {\em cosmic equation of state} 
$p^\CD_{\rm eff} = \beta (\varrho^\CD_{\rm eff}, a_{\cal D})$ with $w_{\rm eff}^\CD
:= p^\CD_{\rm eff}/\varrho^\CD_{\rm eff}$. Now, we may separate the 
{\it morphon equation of state} that plays the role of the {\it Dark Energy
equation of state} \cite{morphon},
\begin{equation}
\label{EOS}
w^\CD_{\Phi} := \frac{{\cal Q}_\CD - 1/3 \langle {\cal R}\rangle_\CD}
{{\cal Q}_\CD + \langle {\cal R}\rangle_\CD}\;.
\end{equation}
We can express the volume state finders through this equation of state parameter and its first time--derivative:
\begin{equation}
\label{statefinder1w}
r^\CD = 1+ \frac{9}{2}w^\CD_{\Phi} ( 1+w^\CD_{\Phi}) (1- \Omega_m^\CD ) - \frac{3}{2}
\frac{{\dot w}^\CD_{\Phi}}{H_\CD} (1-\Omega_m^\CD )\;,
\end{equation}
and
\begin{equation}
\label{statefinder2w}
s^\CD =  1+w^\CD_{\Phi} -\frac{1}{3H_\CD}\frac{{\dot w}^\CD_{\Phi}}{w^\CD_{\Phi}}\;\;.
\end{equation}
As emphasized in \cite{sahnietal,alametal}, the above expressions have the advantage
that one can immediately infer the case of a constant Dark Energy equation of state,
so--called {\it quiessence models}, that here correspond to scaling solutions of the morphon field
with a constant fraction of kinetic to potential energies \cite{morphon}:
\begin{equation}
\label{quiessence}
\frac{2 E^\CD_{\rm kin}}{E^\CD_{\rm pot}} = \frac{\varepsilon{\dot\Phi}_\CD^2 V_\CD}{-U_\CD V_\CD} = 
-1 - \frac{3\CQ_\CD}{\average{\CR}} = 2\frac{w^\CD_{\Phi}+1}{w^\CD_{\Phi} -1}\;\;,
\end{equation}
where the case $\CQ_\CD =0$ (no kinematical backreaction), or $w^\CD_{\Phi} = -1/3$ (i.e. 
$\varrho^\CD_{\Phi} + 3p^\CD_{\Phi} = 0$) corresponds to the `virial condition' 
\begin{equation}
\label{virial}
2\, E^\CD_{\rm kin} \,+\, E^\CD_{\rm pot} =\;0\;\;,
\end{equation} 
obeyed by the scale--dependent Friedmannian model. Again, a non--vanishing backreaction
is associated with violation of `equilibrium'. Note that a morphon field does not violate
energy conditions as in the case of a fundamental scalar field (see \cite{morphon}).
Again it is worth emphasizing that the above--defined equations of state are scale--dependent.

With the help of these dimensionless parameters an inhomogeneous, anisotropic and scale--dependent state can be effectively characterized.

\section{Implications, further insights and prospects}

Having laid down a framework to characterize inhomogeneous cosmologies and having 
understood the physical nature of backreaction effects, does not entitle us to conclude about
the {\em quantitative} importance of inhomogeneities for the global properties of world models.
It may well be that the robustness of the standard model also withstands this challenge.
A good example is provided by Newtonian cosmology that is our starting point for discussing
the implications of the present framework.

\subsection{Thoughts on Newtonian cosmology and N--body simulations}

Analytical as well as numerical models for inhomogeneities are commonly studied within Newtonian cosmology. Essential cornerstones of our understanding of inhomogeneities
rest on the Euclidean notion of space and corresponding Euclidean spatial averages. 

\subsubsection{Global properties of Newtonian models}

The present framework can also be applied to the Newtonian equations and, indeed,
at the beginning of its development the main result on global properties of Newtonian
models was the confirmation of the FRW cosmology as a correct model describing the averaged
inhomogeneous variables. Technically, this result is due to the fact that the averaged principal
invariants, encoded in $\CQ_\CD$, are complete divergences on Euclidean space sections and,
therefore, have to vanish on some scale where we impose periodic boundary conditions
on the deviation fields from the FRW background.  The latter is a necessary requirement
to obtain unique solutions for Newtonian models (for details see \cite{buchertehlers}). 

This point is interesting in itself, because researchers who have setup cosmological
N--body simulations did not investigate backreaction: the vanishing of the averaged 
deviations from a FRW background
was enforced by construction. The same remark applies to analytical models, where a
homogeneous background is introduced with the manifest implication of coinciding with
the averaged model, but without an explicit proof.
The outcome that a FRW cosmology indeed describes the average of a general
Newtonian cosmology correctly can be traced back to the (non--trivial) property that the
second principal invariant $\rm II$ can indeed be written (like the first) as a complete divergence. 
Since this latter fact is not valid in Riemannian geometry, backreaction studies pose 
a clear task for generalizing current cosmological simulations and analytical models. If backreaction
is substantial, then current models must be considered as toy--models that have improved
our understanding of structure formation, but are inapplicable in circumstances where
the dynamics of geometry is a relevant issue. We shall learn that these circumstances are
those needed to route Dark Energy back to inhomogeneities.

\subsubsection{Backreaction views originating from Newtonian cosmology and relativistic
perturbation theory of a FRW background}

As a rule of thumb we can say that any model (including relativistic models) that involves
periodic boundary conditions imposed on fields on Euclidean space must be rejected as a potential candidate for a backreaction--driven cosmology. Such models
are simply too restrictive to account for a non--vanishing (Hubble--scale) $\CQ_\CD$ that is a generic
property of relativistic models. Of course, also in those models, backreaction can be
investigated (a detailed investigation within Newtonian cosmology may be found in \cite{bks}
as well as an application on the abundance statistics of collapsed objects \cite{abundance}), but it is then only a rephrasing of the known {\em cosmic variance}
within the standard model of cosmology. 

We refer to the term `quasi--Newtonian' when we think of models that are
restricted to sit close to a Friedmannian state. Although we do not refer to the discussion of
structure on super--Hubble scales \cite{rasanen:darkenergy}, \cite{kolbetal:superhubble}, \cite{branden1}, \cite{parry}, the following 
consideration would also apply there. The
integrability condition (\ref{integrability}), in essence, spells out the generic 
coupling of kinematical fluctuations to the evolution of the averaged scalar curvature.
If that coupling is {\em absent}, this equation shows that $\CQ_\CD \propto V_\CD^{-2}$ and $\average{\CR}\propto a_D^{-2}$, i.e. the averaged curvature evolves like a constant--curvature model, and backreaction decays more rapidly than the averaged density, $\average{\varrho}\propto V_\CD^{-1}$. In other words, backreaction cannot be
relevant today (we shall make this more precise in the following).
Therefore, as another rule of thumb, we may say that {\em any
(relativistic) model that evolves curvature at or in the vicinity of the constant--curvature
model is rejected as a potential candidate for a backreaction--driven cosmology} \cite{morphon}.   

In summary, {\em there is no Dark Energy that can be
routed back to inhomogeneities on large scales in Newtonian and quasi--Newtonian models}.

\subsection{Qualitative picture for backreaction--driven cosmologies}

If we just look at the definition of averaging, any value of a volume--averaged quantity
is damped by volume expansion. As for the backreaction term $\CQ_\CD$, the relevant 
positive term that could potentially drive an accelerated expansion 
in accord with recent indications from supernovae data \cite{SNLS}\footnote{Note, however, that
the interpretation of volume acceleration in those data relies on the FRW cosmology.},
is the averaged variance of the rate of expansion, {\em cf.} Eq.~(\ref{backreactionterm}). 
This term, however, is quadratic and damped by the square of the inverse volume.
How can we then expect that, in an expanding Universe, such a term can  
be of any relevance at the present time? Before we give an answer to this question, let us make
precise the notion of `relevance'. This can be done with the help of the averaged equations
as has been first advocated by Kolb {\em et al.} \cite{kolbetal,kolbetalc}.

\subsubsection{Acceleration condition}

Let us look at the general acceleration law (\ref{averagedraychaudhuri}),
and ask when we would find {\em volume acceleration} on a given patch of the spatial
hypersurface \cite{kolbetal}, \cite{buchert:darkenergy,buchert:static}:
\begin{equation}
\label{accelerationcondition}
3\frac{{\ddot a}_\CD}{a_\CD} = \Lambda 
-4\pi G \langle\varrho\rangle_\CD + {\cal Q}_\CD \;>\;0\,.
\end{equation}
We find that, if there is no cosmological constant, the necessary condition ${\cal Q}_\CD > 4\pi G\langle\varrho\rangle_\CD$  must be satisfied on a sufficiently large scale, at least at the present time.
This requires that ${\cal Q}_\CD$ is positive, i.e. shear fluctuations are superseded by
expansion fluctuations {\em and}, what is crucial, that ${\cal Q}_\CD$ decays less rapidly than the averaged density \cite{buchert:darkenergy}. 
It is not obvious that this latter condition could be met in view of our remarks
above. We conclude that backreaction has only a chance to be {\em relevant} (e.g. as defined through the inequality Eq.~(\ref{accelerationcondition}) today), if its 
decay rate substantially deviates from its `Newtonian' behavior and, more precisely,
its decay rate must be weaker than that of the averaged density (or at least comparable,
depending on initial data for the magnitude of {\em Early Dark Energy}
\cite{caldwell&wetterich}, \cite{caldwell&linder}). 

\subsubsection{Curvature--fluctuation coupling}

It is clear by now that a {\em backreaction--driven cosmology} \cite{rasanen} must make efficient use
of the genuinly relativistic effect that couples kinematical fluctuations to the averaged
scalar curvature, as is furnished by the integrability condition (\ref{integrability}) (or the
Klein--Gordon equation (\ref{kleingordon}) in the mean field description). 
While the Newtonian and quasi--Newtonian frameworks suppress the scalar field degrees
of freedom attributed to backreaction (or the `morphon field' in the mean field description) by construction \cite{wald},  and so cannot lead to an explanation of Dark Energy on the Hubble scale, general relativity offers a wider range
of possible cosmologies, since it is not constrained by the assumption of Euclidean or
constant--curvature geometry and small deviations thereof. 
But, how can a cosmological model be driven away from a `near--Friedmannian' state, if
we do not already start with initial data away from a perturbed Friedmannian
model? What is the {\em mechanism} of a backreaction--driven cosmology?

\subsubsection{The `Newtonian anchor'}

Let us guide our thoughts by the following intuitive picture.
Integral properties of Newtonian and quasi--Newtonian models
remain unchanged irrespective of whether fluctuations are absent or `turned on'. 
Imagine a ship in a silent water and wind
environment (homogeneous equilibrium state). Newtonian and quasi--Newtonian models do not allow,
by construction, that the ship would move away as soon as  water and wind become more violent.  This `Newtonian achor' is lifted into the ship as soon as we allow for the coupling
of fluctuations to the geometry of spatial hypersurfaces in the form of the averaged
scalar curvature. It is this coupling that can potentially drive the ship away, i.e. change the
integral properties of the cosmology.
Before we are going to exemplify this coupling mechanism, e.g. by discussing  exact solutions, let us add some understanding to the role played by the averaged scalar curvature.

\subsubsection{The role of curvature}

Looking at the integral of the {\em curvature--fluctuation--coupling}, Eq.~(\ref{integrabilityintegral}), we understand that the constant--curvature of the standard
model is specified by the integration constant $k_{\initial\CD}$. This term does not 
play a crucial dynamical role as soon as backreaction is at work. Envisaging a cosmology that
is driven by backreaction, we may as well dismiss this constant altogether. In such a case, the averaged
curvature is {\em dynamically} ruled by the backreaction term and its history.
Given this remark  we must expect that the averaged scalar curvature may experience
changes in the course of evolution (in terms of deviations from constant--curvature), as soon as the structure formation process injects backreaction. (This picture is actually what one needs
to solve the {\em coincidence problem}.) 

This mechanism can be qualitatively understood by studying {\em scaling solutions}, which
impose a direct coupling, $\CQ_\CD \propto \average{\CR}$. (These scaling solutions correspond to {\em quiessence fields}, Eq.~(\ref{quiessence}), and have been thoroughly studied by many people working on quintessence (see \cite{sahnistarobinskii}, 
\cite{copeland:darkenergy} and references therein.)   
In the language of a `morphon field', the mechanism perturbs the `virial equilibrium',
Eq.~(\ref{virial}), such that the potential energy stored in the averaged curvature is
released and injected into an excess of kinetic energy (kinematical backreaction).
In the relevant case of a `Dark Energy scenario' the averaged curvature becomes 
less positive or more negative, respectively, for positive kinematical backreaction.
Thus, in this picture, positive backreaction, capable of mimicking Dark Energy, is feeded
by the global `curvature energy reservoir'.
It is clear that such a mechanism relies on an evolution of curvature that differs from the
evolution of the constant--curvature part of the standard cosmology.
Indeed, as we shall exemplify below, already a deviation term of the form  $\average{\CR} - k_{\initial\CD}a_\CD^{-2} \propto a_\CD^{-3}$ would be {\em quantitatively} sufficient to 
explain the needed amount of Dark Energy today. 

If we start with `near--Friedmannian initial data', and no cosmological constant, then the averaged curvature must be
{\em negative} today and of the order of the value that we would find for a 
void--dominated Universe \cite{morphon}. 
Thus, curvature evolution is key to realizing backreaction effects in the direction that we
would need to explain Dark Energy. 
The difference to the concordance model is essentially that the averaged curvature
changes from an almost negligible value at the CMB epoch to a cosmologically relevant 
negative curvature today. This is one of the direct hints to put backreaction onto the stage
of observational cosmology. 

Let us add two remarks. First, it is not at all evident that a flat Universe 
is necessarily favoured by the data {\em throughout the evolution} \cite{ichikawa}. This latter
analysis has been performed within the framework of the standard model, and it is clear
that in the wider framework discussed here, the problem of interpreting astronomical data
is more involved. Second, it is often said that spatial curvature can only be relevant near Black Holes
and can therefore not be substantial. Here, one mistakenly implies an astrophysical Black Hole,
while the Schwarzschild radius corresponding to the matter content in a Hubble volume is of
the order of the Hubble scale. In essence, a cosmologically relevant curvature contribution is
tiny, but this property is shared by all cosmological sources.

\noindent
(The above qualitative picture is illustrated in detail in R\"as\"anen's review \cite{rasanen}.)

\subsection{Exact solutions for kinematical backreaction}
\label{subsect:exactsolutions}

The following families of exact solutions of the averaged equations are used to illustrate
the mechanism of a backreaction--driven cosmology. Other implications of these examples
are discussed in \cite{buchert:static}.

\subsubsection{A word on the cosmological principle}

We may separate the following classes of solutions into those solutions that respect the cosmological principle and those that do not.
It is therefore worth recalling the assumptions behind the cosmological principle.
There are various `definitions' of this principle in the literature. The most realistic, however,
refers to the existence of a {\em scale of homogeneity}: in standard cosmology we assume
that there exists a scale beyond which all observables do no longer depend on scale. It is beyond
this scale where the standard model is supposed to describe the Universe on average; it is
simply unreasonable to apply this model, even on average, to smaller scales, since the 
standard FRW model has an in--built scale--independence. On the same grounds, isotropy 
can only be expected on the homogeneity scale and not below.
Accepting the existence of this scale has strong implications, one of them being that
cosmological parameters on that scale are {\em reprensentative} for the whole Universe.
If this were not so, and generically we may think of, e.g. a  decay of average
characteristics with scale all the way to the diameter of the Universe, then also backreaction 
would differ outside the domain of homogeneity. That scale is thought to be well below the 
scale of the observable Universe and within our past--lightcone. Therefore, averaging over non--causally connected regions
delivers the same values as those already cumulated up to the homogeneity scale \cite{rasanen}, \cite{buchertcarfora:Q}.

We are now briefly describing some exact solutions, and we mainly have in mind to learn about
the coupling between curvature and fluctuations.

\subsubsection{Kinematical backreaction as  a constant--curvature or a cosmological constant}

Kinematical backreaction terms can model a constant--curvature term as is already evident
from the integrability condition (\ref{integrability}). Also, a cosmological constant need not
be included into the cosmological equations, since 
${\cal Q}_\CD$ can play this role \cite{buchert:jgrg}, \cite{bks}, \cite{rasanen:constraints},
and even provides a {\em constant} exactly, as was shown in \cite{kolbetal} and \cite{buchert:static}. The exact condition can 
be inferred from Eq.~(\ref{averagedraychaudhuri}) and (\ref{generalexpansionlaw}) and reads:
\begin{equation}
\label{lambdacondition}
 \frac{2}{ a_\CD^2 }
\int_{\initial{t}}^t \rmd t'\ {\cal Q}_\CD\frac{\rmd }{\rmd t'} a^2_\CD(t')
\;\equiv\;{\cal Q}_\CD \;\;,
\end{equation}
which implies ${\cal Q}_\CD = {\cal Q}_{\cal D}(t_i) = const.$ as the only possible 
solution. Such a `cosmological constant' installs, however, via
Eq.~(\ref{integrabilityintegral}), a non--vanishing averaged scalar curvature (even for 
$k_{\initial\CD} =0$):
\begin{equation}
\label{lambdacurvature}
\average{\cal R} \;=\;\frac{6 k_{\initial\CD}}{a_\CD^2} \;-\;3 {\cal Q}_\CD (t_i)\;\;.
\end{equation}
This fact has interesting consequences for `morphed' inflationary models \cite{morphedinflation}.

\subsubsection{Out--of--equilibrium Einstein cosmos}

Following Einstein's thought to construct a globally static model, we 
may require the effective scale--factor $a_{\Sigma}$ on a simply--connected 3--manifold $\Sigma$ without boundary to be 
constant on some time--interval, hence ${\dot a}_{\Sigma} = {\ddot a}_{\Sigma}=0$ and
Eqs.~(\ref{averagedraychaudhuri}) and (\ref{averagedhamilton}) 
may be written in the form:
\begin{equation}
\label{static1}
{\cal Q}_{\Sigma}\;=\;4\pi G \frac{M_{\Sigma}}{\initial{V}a_{\Sigma}^3} - \Lambda\;\;\;;\;\;\;
\gaverage{\CR}\;=\; 12\pi G \frac{M_{\Sigma}}{\initial{V}a_{\Sigma}^3}+3\Lambda\;\;,
\end{equation}
with the global {\it kinematical backreaction} ${\cal Q}_{\Sigma}$, the globally averaged scalar 3--Ricci 
curvature $\gaverage{\CR}$, and the total restmass  $M_{\Sigma}$ contained in
$\Sigma$.

Let us now consider the case of a vanishing cosmological constant: $\Lambda = 0$.
The averaged scalar curvature  is, for a non--empty Universe, 
always {\it positive}, and 
the balance conditions (\ref{static1}) replace Einstein's balance 
conditions that determined the cosmological constant in the standard 
homogeneous Einstein cosmos. 
A globally static inhomogeneous cosmos without a cosmological constant 
is conceivable and characterized by the cosmic equation of state:
\begin{equation}
\label{cosmicstate1}
\gaverage{\CR}\;=\;3{\cal Q}_{\Sigma}\;=\;const.\;\;\Rightarrow\;\;
{p}^{\Sigma}_{\rm eff}\;=\; \varrho^{\Sigma}_{\rm eff}\;=\;0\;\;.
\end{equation}
Eq.~(\ref{cosmicstate1}) is a simple example of a strong coupling between curvature and
fluctuations. Note that, in this cosmos, the effective Schwarzschild radius
is larger than the radius of the Universe,
\begin{equation}
\label{blackhole}
a_{\Sigma} = \frac{1}{\sqrt{4\pi G \gaverage{\varrho}}}\;=\;\frac{1}{\pi} 2 G M_{\Sigma}= \frac{1}{\pi} a_{\rm Schwarzschild}\;,
\end{equation}
hence confirms the cosmological relevance of curvature on the global scale $\Sigma$.
The term `out--of--equilibrium' refers to Subsection~\ref{subsect:information}: in the above
example volume expansion cannot compete with information production because the volume
is static, while information is produced (see \cite{buchert:static} for more details).

Such examples of global restrictions imposed on the averaged equations do not refer to 
a specific inhomogeneous metric, but should be thought of in the spirit of the virial theorem
that also specifies integral properties but without a guarantee for the existence of
inhomogeneous solutions that would satisfy this condition.

\subsubsection{A globally stationary inhomogeneous cosmos}

Suppose that the Universe indeed is hovering around 
a non--accelerating state on the largest scales. 
A wider class of models that balances the fluctuations and the 
averaged sources can be constructed by introducing 
{\it globally stationary effective cosmologies}: the vanishing of the second time--derivative
of the scale--factor would only imply ${\dot a}_{\Sigma} = const.=:{\cal C}$, i.e.,
$a_{\Sigma} = a_S + {\cal C}(t-t_i)$, where the integration constant $a_S$
is generically non--zero, e.g. the model may emerge \cite{ellis:emergent1}, \cite{ellis:emergent2}
from a globally static cosmos, $a_S :=1$, or from a  
`Big--Bang', if $a_S$ is set to zero. 
In this respect this cosmos does not appear very different from the standard model, since it evolves at an effective Hubble rate $H_{\Sigma} \propto 1/t$. (There are, however, substantial
differences in the evolution of cosmological parameters, see \cite{buchert:static}.)

The averaged equations
deliver a dynamical coupling relation between  
${\cal Q}_{\Sigma}$ and $\gaverage{\CR}$ as a special case of the integrability condition
(\ref{integrability}):
\begin{equation}
\label{coupling}
-\partial_t {\cal Q}_{\Sigma} + \frac{1}{3} \partial_t  \gaverage{\CR}\;=\;
\frac{4{C}^3}{a_{\Sigma}^3}\;.
\end{equation}
The cosmic equation of state of the $\Lambda-$free stationary cosmos and its solutions read \cite{buchert:darkenergy,buchert:static}:
\begin{eqnarray}
\label{cosmicstate3}
{p}^{\Sigma}_{\rm eff}\;=\; 
-\frac{1}{3}\;\varrho^{\Sigma}_{\rm eff}\;\;\;;\;\;\;
\label{stationaryS}
{\cal Q}_{\Sigma} \;=\; \frac{{\cal Q}_{\Sigma}(t_i)}{a_{\Sigma}^{3}}\;;\\
\label{curvatureevolutionR}
\gaverage{\CR}=  \frac{3 {\cal Q}_{\Sigma}(t_i)}{a_{\Sigma}^3} - 
\frac{3{\cal Q}_{\Sigma}(t_i)-\gaverage{\CR}(t_i)}{a_{\Sigma}^2}\;.
\end{eqnarray}
The total kinematical backreaction ${Q}_{\Sigma}V_{\Sigma} = 4\pi G M_{\Sigma}$ 
is a conserved quantity in this case.

The stationary state tends to the
static state only in the sense that, e.g. in the case of an expanding cosmos, 
the rate of expansion slows down, but the steady increase of the scale factor allows for a global
change of the sign of the averaged scalar curvature.   
As Eq.~(\ref{curvatureevolutionR}) shows, an initially
positive averaged scalar curvature would decrease, and eventually would become 
negative as a result of backreaction.
This may not necessarily be regarded as a signature of  a global topology change, as 
a corresponding sign change in a Friedmannian model would suggest (see Subsect:~\ref{subsect:global}). 

The above two examples of globally non--accelerating universe models evidently violate the cosmological principle, while they would
imply a straightforward explanation of Dark Energy on regional (Hubble) scales:
in the latter example the
averaged scalar curvature has acquired a piece $\propto a_{\Sigma}^{-3}$ that, astonishingly,
had a large impact on the backreaction parameter, changing its decay rate from
$\propto a_{\Sigma}^{-6}$ to $\propto a_{\Sigma}^{-3}$. This is certainly enough to 
produce sufficient `Dark Energy' on some regional patch due to the presence of strong 
regional fluctuations. However, solutions that respect the cosmological principle and, at the
same time, satisfy observational constraints can be
constructed.
As an example, scaling solutions \cite{morphon} have been exploited for such  
a more conservative approach, also outlined in Subsect.~\ref{subsect:outlook}.

\subsubsection{Explicit inhomogeneous solutions}

If we wish to specify the evolution of averaged quantities without resorting to phenomenological assumptions on the equations of state of the various ingredients, or on
the necessarily qualitative analysis of scaling solutions, or with specific global assumptions,
we have to specify the inhomogeneous metric.
Natural first candidates are the spherically--symmetric
Lema\^\i tre--Tolman--Bondi (LTB) solutions that were first employed in the context
of backreaction in \cite{celerier1} and \cite{rasanen:LTB}.

Considerable effort has been spent on LTB solutions and, especially recently, relations to 
integral properties of averaged cosmologies have been sought. Interestingly, 
\cite{nambu} also found a strong coupling between averaged scalar curvature
and kinematical backreaction, and LTB solutions also feature an additional curvature piece  $\propto a_{\CD}^{-3}$ on some domain $\CD$. There are obvious shortcomings of LTB solution studies that consider the class of on average vanishing scalar curvature, since in that
class also $\CQ_\CD \equiv 0$ \cite{singh1};
also here, a non--vanishing averaged curvature is crucial to study backreaction \cite{chuang}.
However, there is enough motivation to study the extra effect of a positive expansion variance
to fit observational data (\cite{celerier2} and references below).

The value of LTB studies is more to be seen in the specification of 
observational properties such as the luminosity distance in an
inhomogeneous metric \cite{LTBgron,LTBluminosity5,LTBluminosity1,LTBluminosity2,LTBluminosity4,alnes2}.
Although interesting results were obtained, especially in connection with the interpretation
of supernova data, care must be taken when determining e.g. just luminosity distances, since
the free LTB functions may fit any data \cite{mustapha}. 
Generally, apart from mistakes (e.g. setting the shear to zero), those studies sometimes
confuse integral properties of a cosmological model with local properties (e.g. the scale factor $a_\CD$ and a local scale factor in the given metric form). The averaged equations cannot
predict luminosity distances unless one considers averages on the lightcone (see, however,
the strategy by Palle \cite{palle}), which in turn
is related to the issue of light--propagation in an inhomogeneous Universe
(see \cite{kantowski}, \cite{ruth:luminosity}, and discussion and references in 
\cite{ellisbuchert}).

\subsection{Where to go from here?}
\label{subsect:outlook}
In this subsection we are going to outline several strategies towards the goal of 
understanding the quantitative importance of backreaction effects, and to device methods of their
observational interpretation. All the topics listed are the subject of  work in progress.

\subsubsection{Global aspects}
\label{subsect:global}

The question of what actually determines the averaged scalar curvature is an open problem.
For a two--dimensional Riemannian manifold this question is answered through the
Gauss--Bonnet theorem: the averaged scalar curvature is determined by the Euler--characteristic
of the manifold. Hence, it is a global topological property rather than a certain restriction
on local properties of fluctuations that determines the averaged scalar curvature. If such an argument would carry over to a three--dimensional manifold, then any local argument for an estimate of backreaction would obviously be off the table. (There are related thoughts and results in
string theory that could be very helpful here.)
In ongoing work \cite{carforabuchert:perelman,buchertcarfora:Q} we consider the consequences of Perelman's proofs that were mentioned
in Subsect.~\ref{subsubsect:ricciflow}. There is no such theorem like that of Gauss and Bonnet
in 3D, but there are uniformization theorems that could provide similar conclusions.
For example, for closed inhomogeneous universe models we can apply Poincar\'e's conjecture (now proven by Perelman) that any simply--connected three--dimensional Riemannian manifold without boundary  is a homeomorph of a 3--sphere. 
Ongoing work concentrates on the multi--scale analysis of the curvature distribution and the related distribution of
kinematical backreaction on cosmological hypersurfaces that feature the phenomenology we
observe. All these studies underline the relevance of topological issues for a full understanding
of backreaction in relativistic cosmology. To keep track with the developments in 
Riemannian geometry and related mathematical fields will be key to advance 
cosmological research. 
In this line it should be stressed that the averaged scalar curvature is only a weak descriptor for the topology in the general 3D case, and information on the sectional curvatures or the full Ricci tensor is required.
In observational cosmology there are already a number of efforts,
e.g. related to the observation of the topological structure of the Universe derived from CMB maps (for further discussion see \cite{buchert:static} and for topology--related issues see 
\cite{uzan:topology,lehoucq:topology}, \cite{cornishspergel}, \cite{aurich1:topology,aurich2:topology}). 

\subsubsection{Approximation methods for backreaction}
\label{subsect:approximations}

There is a large body of possibilities to construct a generic inhomogeneous metric.
First, there is the possibility of using standard methods of perturbation theory.
Although the equations and `parameters' discussed in this lecture can live without introducing
a background spacetime, a concrete model for the backreaction terms can be obtained by
employing perturbation theory (preferably of the Lagrangian type) and, hence, a reference background must be introduced. But, the construction idea is (i) {\em to only model
the fluctuations by perturbation theory (the term $\CQ_\CD$) and to find the final (non--perturbative) model 
by employing the exact framework of the averaged equations}. Such a model is currently 
investigated by paraphrasing the corresponding Newtonian approximation \cite{bks}. 
Second, we could aim at finding an approximate evolution equation for $\CQ_\CD$ 
by (ii) {\em closing the hierarchy of ordinary differential equations} that involve the evolution of shear
and the electric and magnetic parts of the projected Weyl tensor.
In this line, (iii) {\em further studies of cosmic equations of state} (like, e.g., the Chaplygin 
state \cite{chaplygin})  are not only a clearcut way to close the averaged equations, but also a way to classify different solution sectors. 
All these models could be subjected to (iv) {\em standard dynamical system's analysis to show their
stability in the phase space of their parameters} \cite{wainwrightellis,henk:attractor}. Note in this context that the FRW cosmology as an averaged model is found to be stable in many cases,
but there is an unstable sector that just lies in the right corner needed to explain 
`Dark Energy', as was found in  \cite{morphon}.

\subsubsection{Issues of interpretation of backreaction within observational cosmology}

Here, the most important step will be (i) {\em to investigate the averaging formalism on the lightcone}.
Such a framework is currently constructed. It relates not only to all aspects of observations
in terms of distances within inhomogeneous cosmologies, but also links directly to initial
data in the form of, e.g., CMB fluctuation amplitudes and the integrated Sachs--Wolfe effect. Relating lightcone averages to cosmological model averages is also possible and is in the focus
of this investigation. The consequences of a quantitative importance of an integrated backreaction history on the lightcone are obvious. Applying generic redshift--distance
relations e.g. to galaxy surveys would put us in the position to better understand the
actual distribution of galaxies that are currently mapped with the help of FRW distances.  
If expansion fluctuations are a dominant player on large--scales, we can imagine that also
the galaxy density maps would be affected.

Several times we have already pointed out that (ii) {\em scale--dependence of observables}
is key to understand the cosmological parameters in the present framework.
Viewing observational data with this additional discrimination power of a scale--dependent interpretation of backreaction effects, there is furthermore (iii) {\em a link to the Dark Matter problem} that certainly is an important task to be formulated,
and some more words on this particular problem are in order. 

\subsubsection{A common origin of Dark Energy and Dark Matter?}

Concentrating on the Dark Energy problem has led us to focussing on a positive contribution
of ${\cal Q}_\CD$ on large scales. However, the 
kinematical backreaction ${\cal Q}_\CD$ itself can also be negative, and a sign--change may actually happen by going to smaller scales. Looking at the phenomenology
of large--scale structure reveals strongly anisotropic patterns, so that it is not implausible
that on the scales of superclusters of galaxies we would find a shear--dominated ${\cal Q}_\CD < 0$ (this was actually found in the Newtonian investigation \cite{bks} that, however, suffers
from the fact that ${\cal Q}_\CD$ is restricted to drop to zero on the periodicity scale of the fluctuations). Thus, again as a result of its scale--dependence, the kinematical backreaction parameter
can potentially be the origin of a {\em Kinematical Dark Energy}, but also of 
a {\em Kinematical Dark Matter} \cite{buchert:valencia}.

Mapping kinematical backreaction with a `morphon field' opens further links to previous
studies that tried to model Dark Energy {\em and} Dark Matter by a scalar field
(\cite{paddy:darkmatter} and references to earlier work therein).

With this in mind, the volume deceleration functional (\ref{deceleration})
can change sign too, but this crucially depends on the value of the matter density parameter $\Omega^\CD_m$. 
Let us now discriminate different spatial scales. We denote by $\CH$ the homogeneity scale,
by $\CE$ the scale of a typical void, and by $\CM$ the scale of a matter--dominated region
(e.g. the scale of galaxy clusters) \cite{buchertcarfora:Q}.  We notice from Eq.~(\ref{deceleration}) that,
for a small value of $\Omega^\CH_m$, a smaller negative value of
$\Omega^\CH_{\cal Q}$ is needed to obtain volume acceleration on the homogeneity scale, $q^\CH < 0$. The precise criterion for volume acceleration {\it today} is 
$-\Omega^{\now\CH}_{\cal Q} > \Omega^{\now\CH}_m /4$.
Thus, for $\Omega^{\now\CH}_m \approx 1/4$ we would need 
$-\Omega^{\now\CH}_{\cal Q} > 1/16$,
while for a model with no (exotic) Dark Matter, e.g. $\Omega^{\now\CH}_m \approx 1/30$ we would just need  $-\Omega^{\now\CH}_{\cal Q} > 1/120$.
This is a game with numbers until we understand the scale--dependence of the matter
distribution. We know that on smaller scales (from galaxy halos to clusters and superclusters of galaxies) we would need $\Omega^{\now\CM}_m \approx \Omega^{\now\CE}_m \approx 1/4$, and here
there is certainly a problem to explain this amount by baryonic dark matter only. At the moment we can just mention this problem and say that
a quantitative investigation of scale--dependence of the density parameter is an important
task: if no non--baryonic Dark Matter is assumed, it would require
to reconcile a rather large value on smaller scales with a small value on large scales.
In a forthcoming work this issue is further discussed and investigated \cite{buchertcarfora:Q};
it is indeed the case that the matter density parameter drops substantially at around the scale $\CE$ in
a cosmological slice that is volume--dominated by voids.

\subsection{A short conclusion: opening {\em Pandora's Jar}}

Let us conclude by stressing the 
most important issue: {\em quantitative relevance of backreaction effects}. Especially the recent efforts,
spent on the backreaction problem by a fairly large number of researchers,
added substantial {\em qualitative} understanding to the numerious previous efforts that were 
undertaken since George Ellis initiated this discussion in 1984 \cite{ellis} (see references in \cite{ellisbuchert}).
The issue remains unresolved to date: an explanation of Dark Energy along these lines
is attractive and also physically plausible, but a reliable and unambiguous estimate of the actual
influence of these effects is lacking. This situation may change soon and for this to happen it requires
considerable efforts, for which some possible strategies have been outlined in the last subsection.  
 
After those results are coming in, we may face a more challenging situation than anticipated
by the qualitative understanding that we have. For example, while the explanation of Dark Energy by quintessence (or phantom quintessence) still allows to hide the physical
consequences behind a scalar field that is open for a number of explanations, the mapping
of a scalar field to the backreaction problem, as outlined in Subsection~\ref{subsect:morphon}, 
can no longer keep a phenomenological status: {\em fluctuations exist and can be measured}.
There are no free parameters, there are initial data that can be constrained.
Despite being premature, it is nevertheless allowed to speculate that the outcome is
i) a confirmation of the qualitative picture, but ii) a quantitative problem to reconcile
this picture with the data in the sense that there is not enough time for the mechanism
to be sufficient. In that situation we `lost' the standard model for a correct description of the
late Universe, and we do not reach an explanation, unless we allow for initial data that
are non--standard. This in turn asks for a comprehensive understanding of these required
initial data, hence reconsideration of inhomogeneous inflationary models and their
fluctuation spectrum at the exit epoch. We opened Pandora's Jar.

Notwithstanding, I would consider such a situation as the beginning of a fruitful
development of cosmology. As previously mentioned, the issues of scale--dependence
of observables, the priors underlying interpretations of observations, the large subject
of Dark Matter and, of course, the issue of Dark Energy, will be all interlocked and ask for a comprehensive realistic treatment
beyond crude idealizations.  

Even if we would `only' find a $10$ percent effect, rather than $75$ percent, these studies
would have justified their quantitative importance for observational cosmology, and what is
to be expected, would substantially improve our understanding of the Universe. 

\begin{acknowledgments}
I am greatful to M\'ario Novello for his invitation to this school and support
by {\em ICRA}, Brazil, as well as to Roland Triay. Valuable, `deep' thoughts were exhanged with them and with Henri--Hugues Fliche, J.A.S. Lima and Ugo Moschella.

Part of this lecture was written up during a working visit to UCT Cape Town, South Africa,
stimulated by discussions during a lecture course on these issues.
I acknowledge support by the National Research Foundation, South Africa, and 
I am greatful to Peter Dunsby and George Ellis, as well as to Bruce Bassett,
Salvatore Capozziello, Sante Carloni, Chris Clarkson, Charles Hellaby
and Gary Tupper,  for many fruitful conversations and the setup of
interesting collaborations.

I have presented thoughts that I am  sharing with my collaborators Jean--Michel Alimi, Mauro Carfora, J\"urgen Ehlers, George Ellis, Toshi Futamase, Akio Hosoya, Martin Kerscher, Martin Kunz, Julien Larena, 
Masaaki Morita, Aleksandar Raki\'c
and Dominik Schwarz. Thanks go also to Syksy R\"as\"anen and Henk van Elst for constructive comments on the
manuscript, and to
Wolfgang Hillebrandt for his
inspiring remarks on the occasion of the SFB workshop in Gaissach, Germany.
I acknowledge support by the Sonderforschungsbereich SFB 375 
`Astroparticle physics' by the German science foundation DFG.
\end{acknowledgments}

\end{document}